\documentclass[preprint]{aastex63}
\usepackage{fancyhdr}
\usepackage[left]{lineno}
\received{??, 2021}
\revised{??, 2021}
\accepted{\today}
\submitjournal{ApJ}

\shorttitle{Polarization in the Keyhole Nebula}
\shortauthors{Seo et al.}
\begin{document}

\title{Probing Polarization and the Role of Magnetic Fields in Cloud Destruction in the Keyhole Nebula}

\correspondingauthor{Young Min Seo}
\email{youngmin.seo@jpl.nasa.gov}

\author[0000-0003-2122-2617]{Young Min Seo}
\affiliation{Jet Propulsion Laboratory, California Institute of Technology, 4800 Oak Grove Dr. Pasadena, CA, 91109, USA}

\author{C. Darren Dowell}
\affiliation{Jet Propulsion Laboratory, California Institute of Technology, 4800 Oak Grove Dr. Pasadena, CA, 91109, USA}

\author[0000-0002-6622-8396]{Paul F. Goldsmith}
\affiliation{Jet Propulsion Laboratory, California Institute of Technology, 4800 Oak Grove Dr. Pasadena, CA, 91109, USA}

\author{Jorge L. Pineda}
\affiliation{Jet Propulsion Laboratory, California Institute of Technology, 4800 Oak Grove Dr. Pasadena, CA, 91109, USA}

\author[0000-0001-7031-8039]{Liton Majumdar}
\affiliation{National Institute of Science Education and Research, HBNI, Jatni 752050, Odisha, India}

%% Mark off the abstract in the ``abstract'' environment. 
\begin{abstract}

We present polarimetric observations of the Keyhole Nebula in the Carina Nebula Complex carried out using the Stratospheric Observatory for Infrared Astronomy. The Keyhole Nebula located to the west of $\eta$ Carinae is believed to be disturbed by the stellar winds from the star. We observed the Keyhole Nebula at 89 $\mu$m wavelength with the HAWC+ instrument. The observations cover the entire Keyhole Nebula spanning 8$'$ by 5$'$ with central position RA = 10:44:43 and Dec = -59:38:04. The typical uncertainty of polarization measurement is less than 0.5\% in the region with intensity above 5,500 MJy sr$^{-1}$. The polarization has a mean of 2.4\% with a standard deviation of 1.6\% in the region above this intensity, similar to values in other high--mass star--forming regions. The magnetic field orientation in the bar--shaped structure is similar to the large--scale magnetic field orientation. On the other hand, the magnetic field direction in the loop is not aligned with the large--scale magnetic fields but has tight alignment with the loop itself. Analysis of the magnetic field angles and the gas turbulence suggests that the field strength is $\sim$70 $\mu$G in the loop. A simple comparison of the magnetic field tension to the ram pressure of $\eta$ Carinae's stellar wind suggests that the magnetic fields in the Keyhole Nebula are not strong enough to maintain the current structure against the impact of the stellar wind, and that the role of the magnetic field in resisiting stellar feedback in the Keyhole Nebula is limited.

\end{abstract}

%% Keywords should appear after the \end{abstract} command. 
%% See the online documentation for the full list of available subject
%% keywords and the rules for their use.
\keywords{molecular clouds (1072); photodissociation region(1223); Polarimetry (1278); Far infrared astronomy(529)}

%% From the front matter, we move on to the body of the paper.
%% Sections are demarcated by \section and \subsection, respectively.
%% Observe the use of the LaTeX \label
%% command after the \subsection to give a symbolic KEY to the
%% subsection for cross-referencing in a \ref command.
%% You can use LaTeX's \ref and \label commands to keep track of
%% cross-references to sections, equations, tables, and figures.
%% That way, if you change the order of any elements, LaTeX will
%% automatically renumber them.
%%
%% We recommend that authors also use the natbib \citep
%% and \citet commands to identify citations.  The citations are
%% tied to the reference list via symbolic KEYs. The KEY corresponds
%% to the KEY in the \bibitem in the reference list below. 

\section{Introduction} \label{sec:intro}

High-mass star formation is a fundamental process that drives galaxy evolution and the evolution of the interstellar medium (ISM). Recent observations using SOFIA, {\it Herschel}, and radio observatories toward massive star-forming regions - Orion and M17, for example - have highlighted the complex physical processes in these regions  \citep[e.g.,][]{langer14,pabst17,seo19}. However, magnetic fields, which may play a critical role in the evolution of molecular clouds and triggered star formation \citep[e.g.,][]{henney09,mackey11}, have been only sparsely observed. Magnetic fields are incompletely understood since most of these are at large distances from us. In particular, the role of magnetic fields in extreme star--forming regions, which are the closest local analogs of the active star--forming/starburst period in disk galaxies, has been even less-probed than nearby low-mass star-forming regions. Understanding the role of magnetic fields in an extreme star-forming region would provide a critical window to probe how magnetic fields regulate the ISM life cycle in galaxies during starburst periods.     

Only in a handful of high--mass star--forming regions – e.g., Orion, DR21, IC1396, and the Eagle Nebula - has the magnetic field been thoroughly observed. The data show that magnetic fields in such regions are typically ordered (perpendicular or parallel) to the morphologies of bright-rimmed clouds, bars, pillars, or cometary globules \citep[e.g.,][]{neha16,ching17,ching18,pattle18,soam18}. However, existing observations also show a more complex structure than those in low-mass star-forming regions. For example, in the pillars of the Eagle Nebula, the magnetic field is aligned along the pillars while surrounding magnetic fields are perpendicular to the pillars, which is the opposite trend to what is observed in low-mass star-forming regions \citep{pattle18}. \citet{pattle18} suggested that this may be a result of the interaction of the ionization shock with the dense gas. Observations of massive star-forming regions demonstrate that magnetic fields can be altered by the effect of stellar feedback on molecular clouds \citep[e.g.,][]{ferland09,pattle18,eswaraiah20}. However, the role of magnetic fields in resisting stellar feedback was not investigated, leaving an incomplete picture of the evolution of such regions.  

Observations to date are relatively limited in terms of drawing a full picture of the magnetic field evolution in high-mass star-forming regions, as they either cover only small high--mass star-forming regions, where effects of feedback are limited, or extremely large star--forming regions at relatively low spatial resolution (e.g., 30 Doradus at a spatial resolution of 1.1 pc to 5 pc, \citealp{gordon18}). Details of physical processes, particularly on a scale $<$0.3 pc (a typical globule size) related to magnetic fields in extreme environments (e.g., X-ray dominated regions, violent outbursts, etc.), are still mostly unknown. 

Theoretical studies have shown that stellar feedback may alter the orientation and strength of the magnetic field,  affecting subsequent star formation and destruction of molecular clouds \citep[e.g.,][]{henney09,mackey11}. While these studies have shown the evolution of the magnetic field in irradiated globules, they consider only moderate radiative heating, while other types of feedback such as stellar winds, X-rays, and radiation pressure, are neglected. Therefore, a necessary next step is to investigate whether or not these theories can be extended to extreme star-forming regions (e.g., 30 Doradus or the Carina Nebula Complex).

The Keyhole Nebula in the Carina Nebula Complex (CNC) is an excellent testbed in which to study magnetic field evolution in a cloud under the influence of extreme stellar feedback. The CNC is the most energetic star--forming region in our Milky Way, having ~70 O-type stars \citep{smith06}.  It is also frequently compared to 30 Doradus, an extreme star--forming region in the Large Magellanic Cloud. Within the CNC star--forming region, the Keyhole Nebula is only 1.3 pc (2$'$ at a distance of 2.3 kpc) away from $\eta$ Carinae, and is one of the molecular clouds most severely affected by outbursts and X-rays. For example, there is a loop, which may have been formed by the powerful bipolar winds from $\eta$ Carinae that deformed the surrounding material \citep{smith02}. A high-resolution Hubble image reveals detailed structures within the loop showing bundles of filaments and their orientations  \citep{smith10}. It would be highly significant to see whether or not the magnetic field is parallel to the loop, as is the case for coronal loops and low-density molecular filaments. The Keyhole Nebula is the best region in which to probe how the magnetic field evolves during the deformation of a molecular cloud by stellar winds at a high spatial resolution (specifically the 0.09 pc at $\lambda$ = 89 $\mu$m that can be obtained with HAWC+ polarimeter). The properties of the magnetic field (strength and orientation) in the Keyhole Nebula will serve as strong constraints on hydrodynamics modeling and on understanding the destruction of molecular clouds and triggered star formation in extreme star-forming regions.

The goal of this study is to investigate in detail the effect of the magnetic fields on the ISM and cloud evolution under conditions of strong stellar feedback. We found that the magnetic fields have a strong alignment relative to the local structures shaped by the stellar winds so may be being dragged along with the flowing gas. However, in the loop of the Keyhole Nebula, we found that the pressure exerted by the magnetic field tension appears to be negligible compared to the ram pressure of the stellar wind, suggesting that the magnetic fields play only a modest role in maintaining the loop and acting against stellar feedback.

We elaborate our observations and data reduction technique that we have employed in \S2. In \S3, we describe structures in the Keyhole Nebula and magnetic field vectors obtained from polarization observation using SOFIA HAWC+, and analyze the magnetic field orientation to the gas structures. In \S4, we estimate the magnetic field strength in the loop, compare it to the ram pressure of stellar winds produced by  $\eta$ Carinae, and discuss the role of the magnetic field in the context of stellar feedback. Finally, we summarize our results and discussions in \S5.  

\begin{figure*}
\centering
\includegraphics[angle=0,scale=0.78]{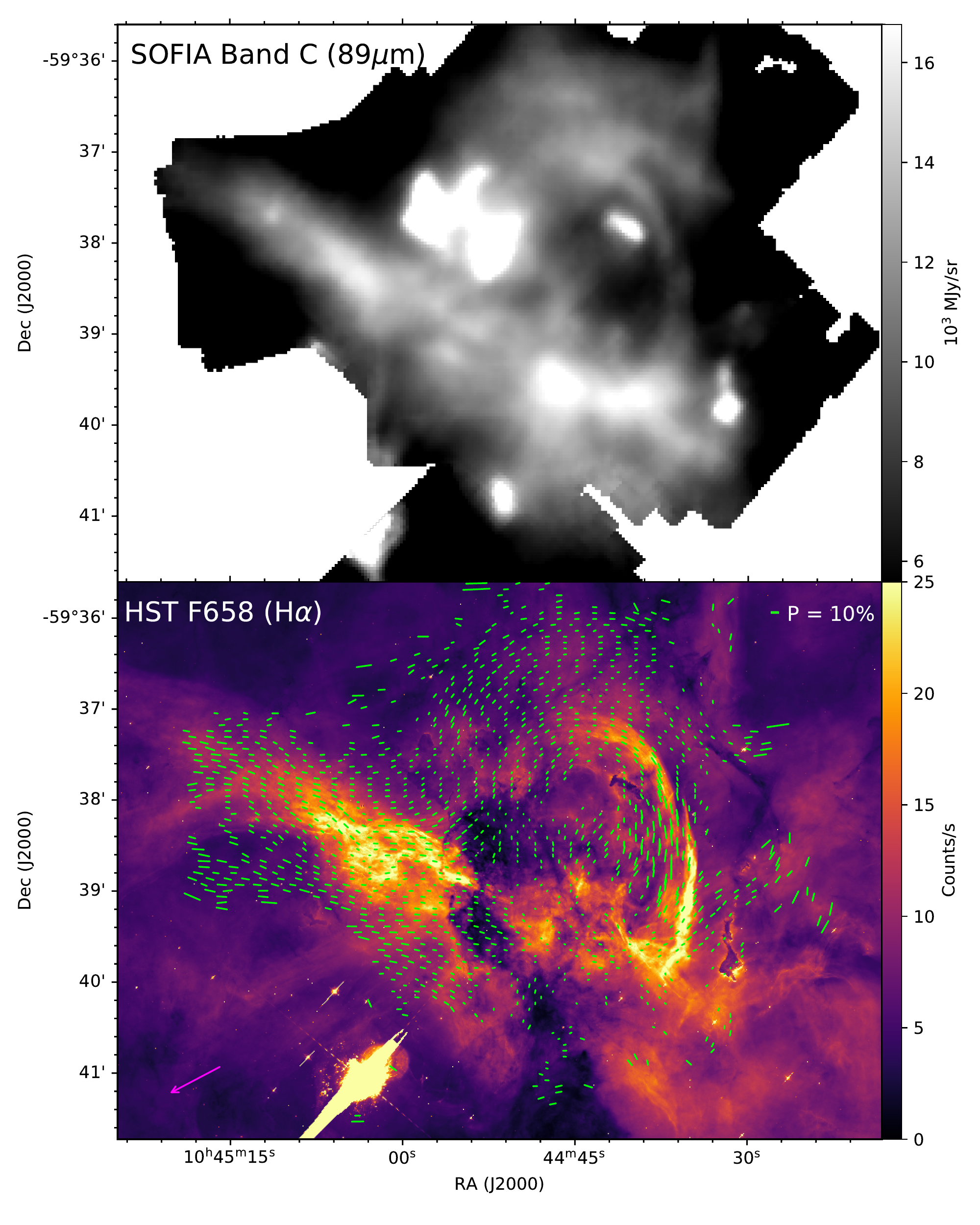}
\caption{Dust continuum image at 89 $\mu$m wavelength obtained using the SOFIA HAWC+ instrument (top) and the H$\alpha$ map of the Keyhole Nebula taken by {\it Hubble} in colored image (bottom). The projected magnetic field vectors are overlaid on the H$\alpha$ image in bright green. The length of a vector with 10\% polarization is indicated at the top right of the bottom panel. Over 2000 vectors were obtained originally in this region, but displaying them all would make the image overly crowded, so in this figure, we only show quarter of the vectors to provide a clearer view of the magnetic field. Linear artifacts are seen extending from the exceptionally bright source $\eta$ Carinae in both intensity images.}
\label{fig_int_continuum}
\end{figure*}

\section{Observations and Data Reduction} \label{sec:obs}

We observed the Keyhole Nebula and $\eta$ Carinae using the HAWC+ instrument on board the Stratospheric Observatory For Infrared Astronomy (SOFIA, \citealp{young12}) as a part of the Guest Observer Cycle 7 campaign (Proposal ID 07$\_$0081, PI: C. D. Dowell). HAWC+ \citep{harper18} was configured to observe at Band C ($\lambda_C$ = 89 $\mu$m).

The Keyhole Nebula is a relatively large structure spanning 15 arcminutes (10 pc at 2.3 kpc distance), which cannot be covered in a single Band C field of view. In consequence, tiled observation areas were determined using the {\it Hubble} H$\alpha$ observation \citep{smith10} and the HI 21 cm map \citep{rebolledo17}. These two observations were used to determine the area to be observed because they show the structures in the Keyhole Nebula most clearly. The observations were made over four flights in July 2018. We used the dithered Nod-Match-Chop mode, which is the standard polarization observation mode for HAWC+. The chop throw (350$''$ -- 500$''$) and direction (within 35$^\circ$ of north-south) were chosen for each tile to avoid bright emission in the chop reference beam. Observations of $\eta$ Carinae and the Homunculus Nebula were made at 53 $\mu$m during the same flight series, and the results will be reported in a separate paper.

The data processing started with the Level 3 products from the DCS pipeline versions 2.3.1 and 2.4.0. Approximately half of the data were affected by a systematic coordinate error of $\sim 33''$, and for each of the two halves the coordinates were adjusted to align with the 70 $\mu$m image from {\it Herschel}/PACS\footnote{For this paper, we used the 70 $\mu$m image generated from observation IDs 1342211615 and 1342211616, downloaded from the Herschel Science Archive (version SPG v14.2.0, JSMAP product, with 2016 July creation date).  Herschel is an ESA space observatory with science instruments provided by European-led Principal Investigator consortia and with important participation from NASA.} \citep{Pilbratt2010, Poglitsch2010} within 3$''$.  The emission in the vicinity of the Keyhole Nebula extends well beyond the $\sim 8'$ chop throw; despite the care in the chop settings, some observations were clearly affected by emission in the reference beam.  To lessen the effects from the reference beam emission, we excluded from each dither position the pixels for which the average emission in the reference beams was greater than 3,300 MJy/sr in the PACS 70 $\mu$m  image. In some cases, the potential holes in the combined image caused by excluding data were filled in with observations having a different chop setting.

The modified Level 3 data were combined with the standard gaussian-- and uncertainty--weighted interpolation in the pipeline, resulting in an angular resolution of 8.4$''$ FWHM (0.09 pc at 2.3 kpc distance).  We applied outlier rejection to the Stokes I, Q, and U images based on redundant observations \citep{Chuss2019}.  Based on a $\chi^2$ test \citep{Santos2019}, we increased the uncertainties in the output images by $\sim$25\% .

The main data products were the fractional polarization (P), position angle (PA), and their uncertainties ($\sigma_P$, $\sigma_{PA}$). The final intensity maps of the Keyhole Nebula and $\eta$ Carinae are shown in Figure \ref{fig_int_continuum} along with the projected magnetic field vectors. The projected magnetic field vectors are derived by rotating the polarization vectors by 90$^\circ$. This is based on the FIR emission of dust grains being polarized parallel to the major axis of dust grains, and the assumption that the major axis of dust grains is aligned at 90$^\circ$ to the magnetic field direction due to the radiative torque \citep[e.g.,][]{draine09}.  The typical statistical polarization uncertainty is less than 0.5\% for the region with Stokes I above 5,500 MJy sr$^{-1}$.    

\begin{figure*}
\centering
\includegraphics[angle=0,scale=0.88]{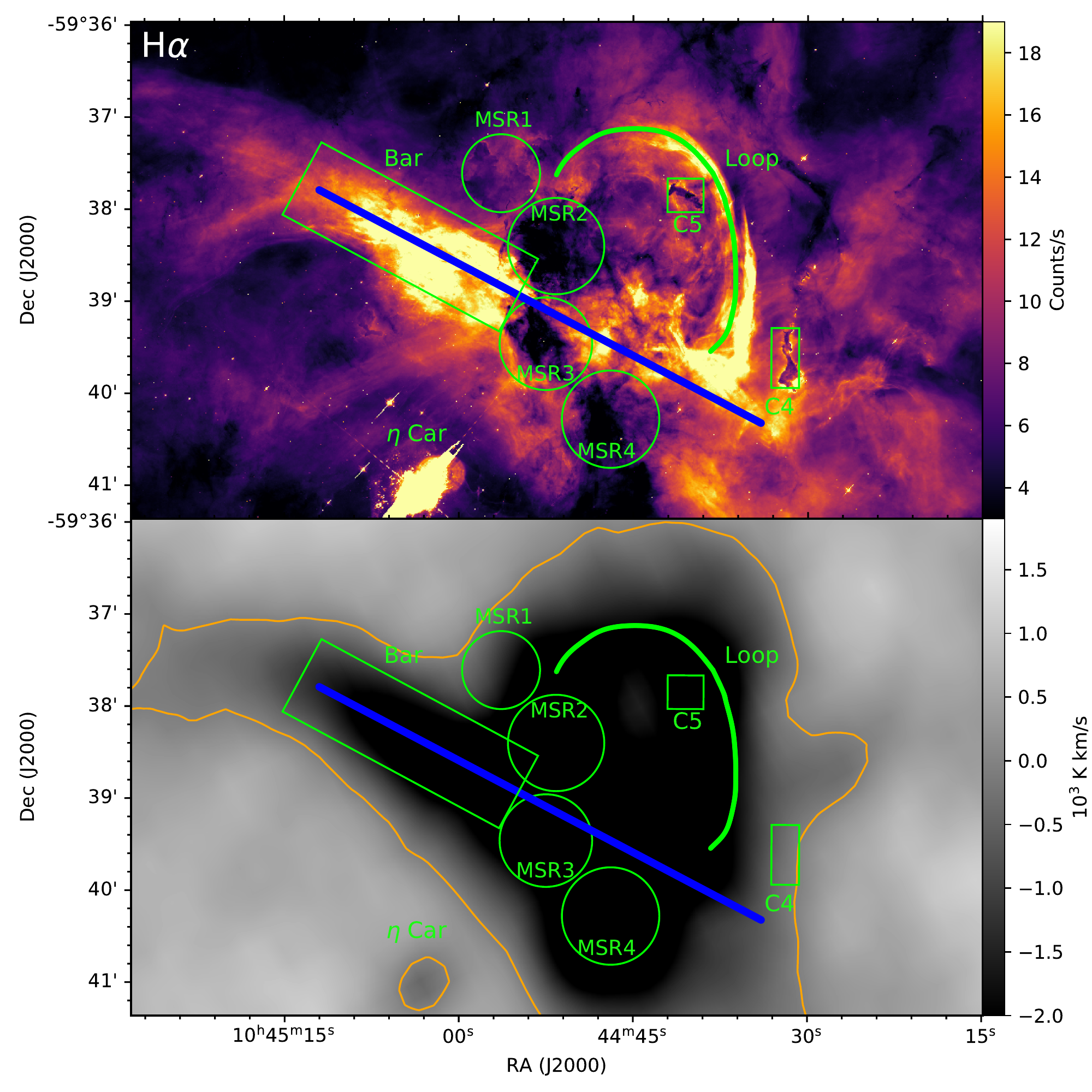}
\caption{H$\alpha$ image of the Keyhole Nebula with the structures on which this study focuses indicated in green (top) and HI 21 cm image (bottom). MSR denotes a multiple structure region since those regions have multiple structures along the line of sight. Clumps 4 and 5 are the largest globulettes in the CNC. The orange contour indicates 0 K km s$^{-1}$ level in HI 21 cm. The region inside the orange contour has negative intensity indicating hydrogen gas in the Keyhole Nebula absorbs the background continuum emission originating from the Tr 16 HII region.} 
\label{fig_structures}
\end{figure*}

\section{Results} \label{sec:mag}
\subsection{Structures in Keyhole Nebula}
We show the Keyhole Nebula at different continuum wavelengths and emission lines in Figures \ref{fig_int_continuum} \& \ref{fig_structures} to describe the structural features in the region. Figure \ref{fig_int_continuum} shows the Keyhole Nebula seen in the 89 $\mu$m Band of the SOFIA HAWC+ instrument, which traces relatively warm dust. The H$\alpha$ image obtained using {\it Hubble} \citep{smith10}, which traces ionized hydrogen, is included in both figures. The H$\alpha$ brightness is proportional to the emission measure; it thus highlights the dense ionization front in this image. The H$\alpha$ image has roughly two orders of magnitude better angular resolution than that of HAWC+ 89 $\mu$m Band, so it delivers a more detailed view of the region. In Figure \ref{fig_structures}, we also show the HI 21 cm image \citep{rebolledo17}. The HI 21 cm emission traces neutral atomic hydrogen and generally indicates the spatial distribution of the warm and cold neutral media whereas $^{12}$CO 1$-$0 traces cold, dense molecular gas. We use an image of $^{12}$CO 1$-$0 in \citet{cox95}. The CO image in \citet{cox95} is the most sensitive CO image toward the Keyhole Nebula, being more sensitive than Mopra CO map \citep{rebolledo16} for this particular region. Below we elaborate overall structures in the Keyhole Nebula visible in the H$\alpha$, CO, and 89 $\mu$m images.

With three different observations in Figures \ref{fig_int_continuum}, \ref{fig_structures}, and CO images in \citet{cox95}, we found that the Keyhole Nebula may be representative of an atypical phase of the ISM. The entire Keyhole Nebula appears in absorption in HI 21 cm (the region within the orange contour in Figure \ref{fig_structures}), and the majority of the region has no significant $^{12}$CO 1$-$0 emission ($<$2 K km s$^{-1}$) except the dark clouds and globulettes (small globules) shown in the H$\alpha$ image \citep[also see ][]{cox95}. This indicates that the Keyhole Nebula does not have a significant amount of dense and cold molecular gas, but there is a large amount of cold atomic hydrogen gas seen against the radio continuum emission background produced by the HII region \citep{brooks01,rebolledo21} This suggests that the Keyhole nebula may be either primarily an atomic cloud, a CO-dark molecular cloud or a molecular cloud transitioning to an atomic cloud. \citet{brooks00} observed this region in H$_2$ 1$-$0 (S1) line and reported no significant emission in the Keyhole Nebula except toward the dark clouds and the globulettes. \citet{roccatagliata13} showed that the dust temperature is around 40 K, suggesting again that the Keyhole Nebula is more likely an atomic than a molecular cloud. This can be further confirmed or shown otherwise with  [C\,{\sc ii}] observations \citep[e.g.,][]{langer14}, which are planned for a Cycle 9 SOFIA Guest Observer program.

This study is focused on two structures: the bar and the loop. While the Keyhole Nebula consists of highly complicated structures and there are many more structures than these two, we use these two structures to probe the role of the magnetic field in resisting stellar feedback. We also probe additional four regions involving dark spots near the bar and the loop in the H$\alpha$ images (Figure \ref{fig_structures}). The dark spots arise due to foreground clouds to the HII region. We probe the four regions to see if the foreground clouds modify polarization emission from the bar and the loop.   

The bar is the filamentary structure indicated by the solid blue line in Figure \ref{fig_structures}. The loop is indicated by the bright green line in the same figure. The bar and the loop can be seen in the 89 $\mu$m dust continuum, H$\alpha$, and HI 21 cm images but not in $^{12}$CO 1$-$0. 

The four regions having dark spots near the bar and the loop are indicated by the green circles (MSR1, MSR2, MSR3, and MSR4 where MSR denotes multiple structure region). In particular, MSR2, MSR3, and MSR4 appear as very dark against the bright H$\alpha$ background coming from the HII region of Trumpler 16 (Tr 16), while MSR1 is slightly darker than the background continuum, suggesting that the cloud in MSR1 may be inside the Tr 16 HII region while the other regions may include cold clouds at the nearside boundary of the Tr 16 HII region. However, the cloud in MSR1 may not be at the same distance as the bar and the loop that is interacting with $\eta$ Carinae, since the photodissociation region of the cloud is brighter at its west surface, which faces the Trumpler 14 (Tr 14) cluster rather than $\eta$ Carinae. Also, the centroid velocity of $^{12}$CO 1$-$0 in MSR1 is -18 km s$^{-1}$ \citep[denoted A1 in ][]{cox95}, while the centroid velocities of the HI 21 cm line of the bar and the loop are about -30 km s$^{-1}$ \citep{rebolledo17}. Within MSR2 and MSR3, there is also $^{12}$CO 1$-$0 emission These regions are indicated as A2 \& 1, and 2, respectively, in \citet{cox95}, suggesting that there are foreground molecular clouds to MSR2 and MSR3. MSR4 does not have any noticeable CO emission, which suggests there may be a foreground atomic or CO-dark cloud. To the south of MSR4, there is also a dark silhouette in the H$\alpha$ image which is the tip of a larger cloud extending South East part of the Tr 16 region \citep{smith10}.  

In the 89 $\mu$m dust continuum, MSR1 and MSR2 have exceptionally bright 89 $\mu$m emission, indicating that there are warm dust grains efficiently heated by the radiation from high-mass stars and that the clouds in MSR1 and MSR2 may be relatively denser than the clouds in MSR3 and MSR4. Indeed, the clouds in MSR1 and MSR2 have bright $^{12}$CO 1$-$0 emission ($>$30 K km s$^{-1}$), suggesting that they are relatively dense molecular clouds \citep{cox95}. On the other hand, MSR3 and MSR4 do not correlate with any significant emission in 89 $\mu$m, but show similar 89 $\mu$m intensity to that of the bar. This suggests either that the clouds in MSR3 and MSR4 may be in a different radiation environment to those in MSR1 and MSR2, or that the column density of the clouds may be not high enough to add significant dust continuum at 89 $\mu$m to the emission from the other structures along the line of sight, the bar. Resolving such details requires further observations and is beyond the scope of the current study.  

There are two relatively-large globulettes (C4 and C5) in the Keyhole Nebula region, which are designated as Clumps 4 and 5 in \citet{cox95} and \citet{brooks00}. They appear as dark structures in H$\alpha$ and are relatively bright in 89 $\mu$m, $^{12}$CO 1$-$0, PAH emission, and H$_2$ 1$-$0 S(1) \citep{cox95,brooks00}. This suggests that they are dense and heated by radiation. However, the bright rim of Clump 4 faces south where there are a few OB stars and the bright rim of Clump 5 faces west toward Tr 14, suggesting that they may not be near to $\eta$ Carinae, the loop, and the bar of the Keyhole Nebula.

\begin{figure*}
% FIGURE 3
\centering
\includegraphics[angle=0,scale=0.88]{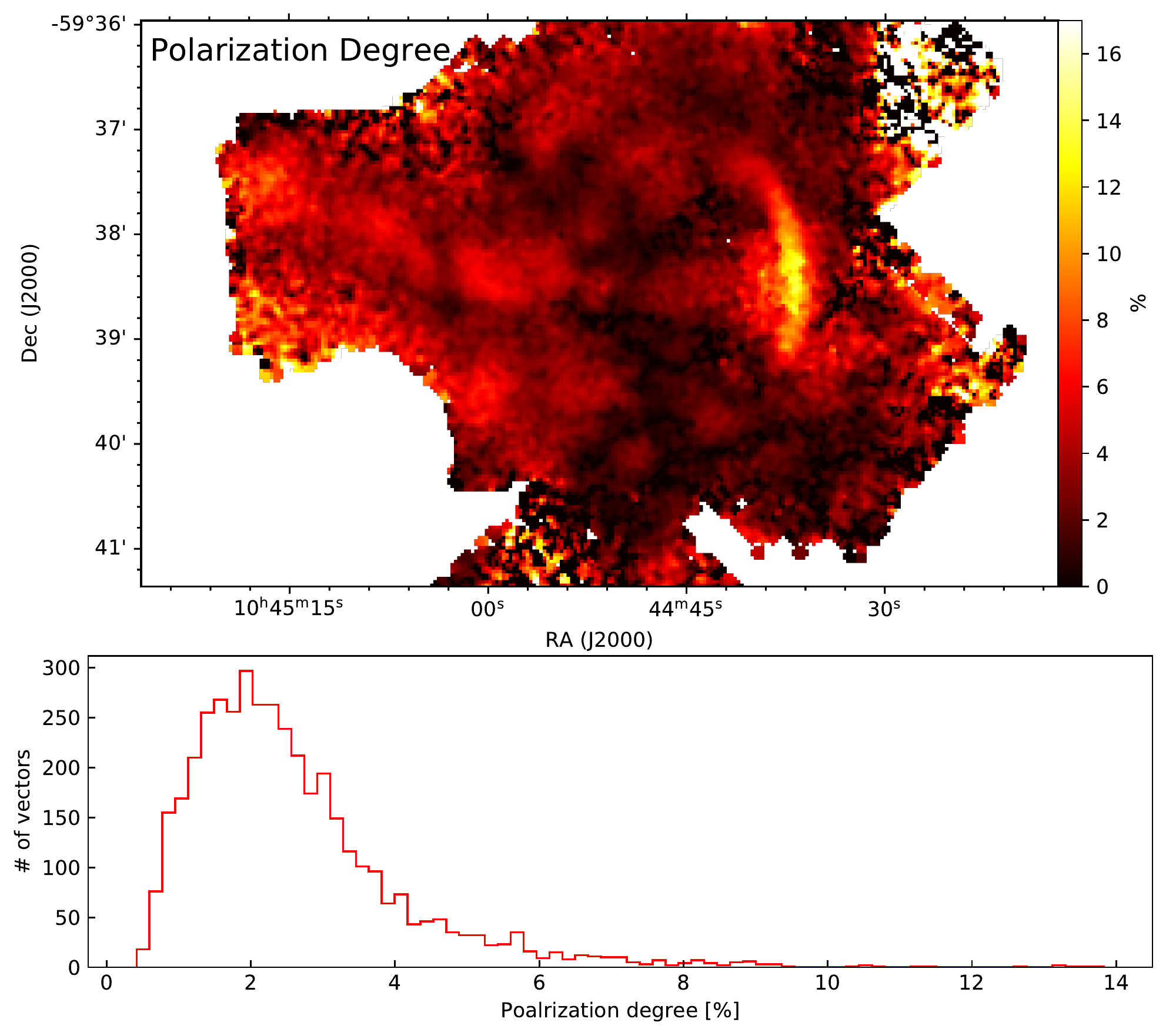}
\caption{Polarization degree image (top) and distributions of polarization degree (bottom) in the Keyhole Nebula. The polarization degree is estimated for vectors in regions of intensity above 5,500 MJy sr$^{-1}$.}
\label{fig_B_total}
\end{figure*}

\begin{figure*}
% FIGURE 4
\centering
\includegraphics[angle=0,scale=0.88]{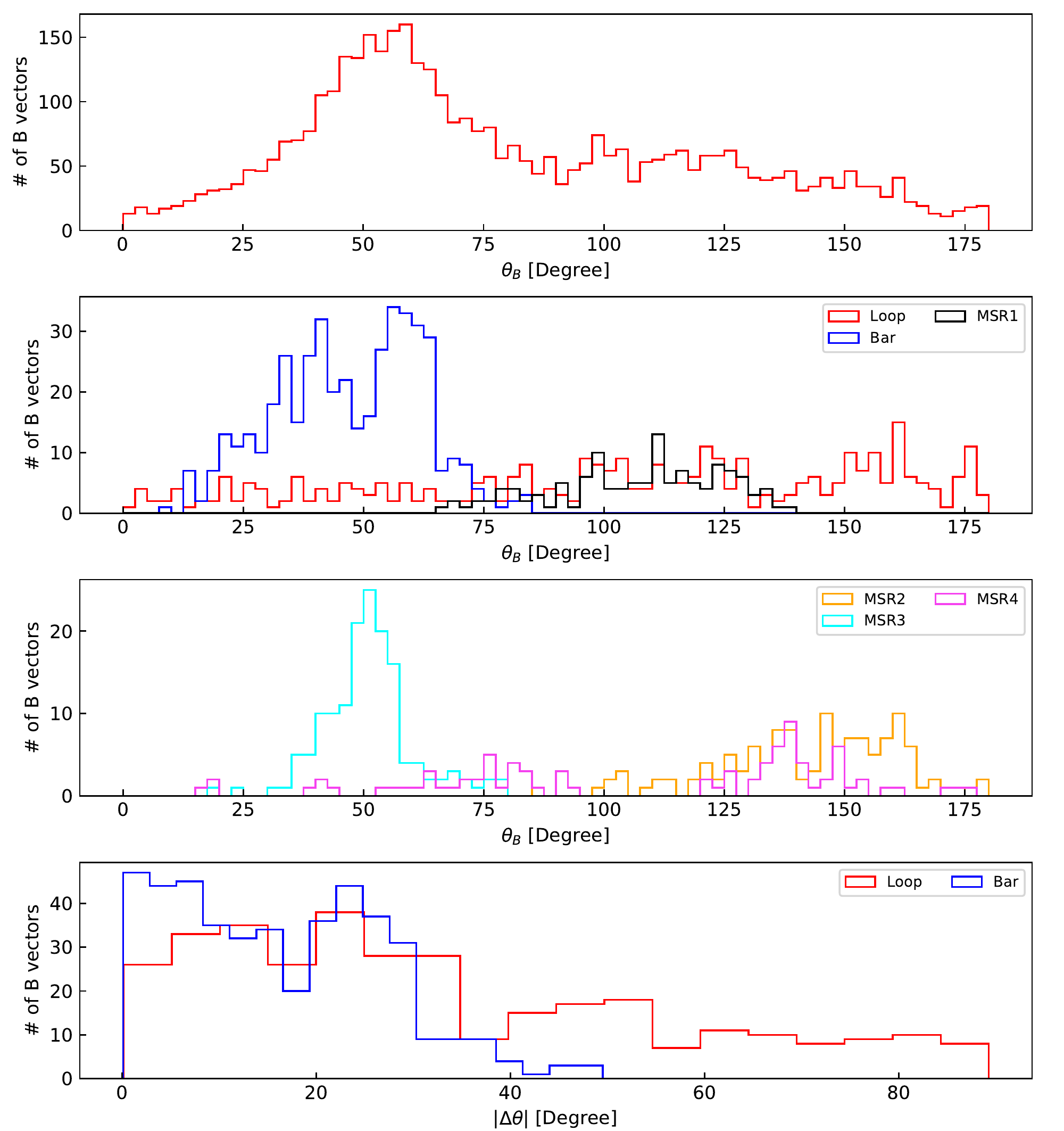}
\caption{Distributions of the angles of the magnetic fields in the regions of intensity above 5,500 MJy/sr (top) and toward the six regions defined in Figure \ref{fig_structures} (2nd and 3rd panels), and the magnetic field orientations with respect to the long-axis of the bar and to the tangent of the loop (bottom). The angle of the magnetic field vector is measured clockwise from East in Galactic coordinates. Different structures have been plotted using different colors as noted at the top-right of each panel. In the bottom panel, the directions of the magnetic fields within the green rectangle in Figure \ref{fig_structures} (blue histogram) are compared to the direction of the heavy blue line in that Figure representing the orientation of the bar. For the loop, we used vectors within a distance of 10$''$ from the solid green line, and we estimated the magnetic field orientations with respect to the tangent of the nearest point on the loop.    }
\label{fig_B_angles}
\end{figure*}

\subsection{Polarization and Magnetic Fields}
The magnetic field vectors are overlaid on the H$\alpha$ image in Figure \ref{fig_int_continuum}, and Figure \ref{fig_B_total} shows a map and histogram of the polarization degree. The mean polarization degree in the region is 2.4 \% with a standard deviation of 1.6 \%.  These values are similar to those in OMC-1 and M17 \citep{vaillancourt12,zeng13}, which have relatively weaker star-formation activity and stellar feedback than the CNC.

We estimated polarization degree in the bar, the loop, and the four MSRs (green circles in Figure \ref{fig_structures}). The polarization degree in the loop is 3.4 \% with a standard deviation of 2.1 \%, while the degree in the bar is 2.6 \% with a standard deviation of 0.81 \%. The maximum polarization degree in the loop is 13.7 \% at the West side of the loop, while the maximum degree in the bar is 4.7 \%. Within the scope of this study, it is hard to say what causes such difference in the polarization degree between the two structures, but our measurements suggest that the dust grains in the loop may be more tightly coupled to the magnetic fields there compared to the the dust grains in the bar.

The four MSRs have an average polarization degree of 1.3 \%  with standard deviation less than 1 \%, except MSR3 that has a mean value of 2.1 \% and a standard deviation of 0.5 \%. These values are significantly lower than those of the loop and the bar. We also found that two globulettes, C4 and C5 have polarization degree in the range of 0.2 -- 1.5\%, which is even lower than polarization degree in the MSRs. These results agree with previous studies showing that denser regions tend to have a lower polarization degree compared to more diffuse regions \citep[e.g.,][]{planck15}.     

The top panel of Figure \ref{fig_B_angles} shows the angle distribution of the entire set of magnetic field vectors measured in the Keyhole Nebula. The angle of each vector is measured clockwise from East in Galactic coordinates. The distribution of magnetic field angles has an interesting form, with a quite sharp peak between 50 and 60 degrees, and a second, much broader component, centered near 90 degrees. In the BLASTPol observations of the Carina Nebula Complex \citep{shariff19}, the large-scale magnetic fields crossing the Keyhole Nebula have angles around 60 degrees, which is near the sharp peak in the figure. However, the vectors with angles from 100 degrees to 180 degrees, which includes roughly 35\% of the total number of vectors, appear to deviate from the large-scale magnetic fields in the region, suggesting that the magnetic fields may have been locally disturbed. 

To further investigate whether or not the magnetic fields are related to the local structures, we analyze magnetic field angles in the structures defined in Figure \ref{fig_structures}. The middle panel of Figure \ref{fig_B_angles} shows the angle distribution of each structure. The analysis of magnetic fields in the bar is done using the vectors only within the green rectangle in Figure \ref{fig_structures}, although the bar is a factor of 2 larger than the longer side of the box as indicated by the blue thick line in the figure. We made this restriction to minimize the contamination by the foreground clouds found in the lower portion of the bar when analyzing  the structure of the magnetic field. For the loop, we analyze the vectors within an angular distance of 5$''$ of the thick green line in Figure \ref{fig_structures}.  This is half of the thickness of the loop, and avoids the MSR1 and MSR2 that include foreground clouds. For the four MSRs, we analyze the vectors within the green circles.    

The angle distribution in the bar has a peaked distribution with median value of 48 degrees and a standard deviation of 15 degrees. The median angle is similar to the angle of the large--scale magnetic field in the Tr 16 region \citep{shariff19}, suggesting that the magnetic field in the bar is aligned with the large-scale structure. The magnetic field in the vicinity of the bar is closely aligned with its major axis, mostly within $\leq$20 degrees (bottom panel of Figure \ref{fig_B_angles}). 

A general finding in polarization studies toward star--forming regions is that the long-axis of filamentary structures is aligned with the local magnetic field direction for low column density clouds (A$_V$ $\leq$3 mag, \citealp{sugitani2011, palmeirim13,franco15, santos16, soler19}). Considering the fact that the Keyhole Nebula may be either a CO-dark molecular cloud or a HI cloud, the extinction is expected to be modest (e.g., A$_V$ $<$ 5, \citealp{bohlin78,tang16}). While we do not have direct measurement of A$_V$, it can be estimated from the hydrogen column density \citep[e.g.,][]{guver09}. \citet{preibisch12} estimated the hydrogen column density in the CNC using dust continuum maps seen by {\it Herschel}. The column density toward the Keyhole Nebula ranges from 1 $\times$ 10$^{21}$ cm$^{-2}$ to 8 $\times$ 10$^{21}$ cm$^{-2}$, which corresponds to A$_v$ = 0.5 $-$ 4. Thus, the alignment of magnetic field direction with the long axis of the bar and the large-scale magnetic fields seems to agree with the general findings in other studies.

The  direction of the magnetic field in the loop spans a wide range, from 0 degrees to 180 degrees, and is almost uniformly distributed. This indicates that the magnetic field in the loop does not follow the large-scale magnetic field. While the angle distribution is almost uniform, the magnetic field angles are not randomly distributed but are aligned with the structure of the loop itself (bottom panel of Figure \ref{fig_B_angles}). 

We have investigated the angle difference between the local tangent to the loop and the magnetic field orientation, and find that half of the vectors are aligned with the loop to within 25 degrees. This may appear as a slightly weaker alignment compared to that in the bar. However, the slightly loose alignment is due in part because the thick green line in Figure \ref{fig_structures} represents the global morphology of the loop, which may consist of an ensemble of small-scale features. These include, for example, intertwined filaments that make wavy deviations from the global morphology of the loop. When considered closely, the magnetic fields are more tightly correlated with the filaments than with the global curvature of the loop. We conclude that on a larger scale, the magnetic field is well aligned with the loop. 

The orientation of the magnetic field between the bar and the loop (the fan region) is unique. The field in this fan region is neither aligned with the loop nor with the bar but rather shows a fan shape starting from the bar and extending to the loop. While this does not prove that the loop formation is induced by the $\eta$ Carinae's stellar wind, the magnetic field orientation agrees with the direction of the gas flow of the stellar wind, similar to the trend that has been often observed in low-density regions in other star-forming regions \citep[e.g., striations in Taurus, ][]{palmeirim13}. This agrees with a scenario that the magnetic field has been disturbed by the stellar wind.   

The magnetic field in MSR1, MSR2, and MSR3 does not show any abrupt disconnection in the orientation angle from those in the loop and the bar. MSR1 and MSR2 contain bright sources in the 89 $\mu$m image (Figure \ref{fig_int_continuum}) where we obtained polarization, but the magnetic fields smoothly connect from the loop and the fan regions to MSR1 and MSR2 without any sudden distortion in field orientation. This suggests either that the polarization in the loop and the fan dominate the polarization along the line of sight or that the magnetic field direction in the foreground clouds happens to be in a similar direction to the background structures. Resolving the two possibilities is beyond the scope of this study since it is practically impossible to do analysis separately on each structure using the dust continuum. The dark cloud in MRS3 is in front of the bar and is not a bright source in the 89 $\mu$m image. The image shows that the field orientation of the bar seems to be continuous across MSR3; thus, the magnetic field orientation in MSR3 could be primarily contributed by the bar. However, we have excluded MSR3 from the polarization analysis of the bar to avoid possible confusion. 

The magnetic field in MSR4 is almost perpendicular to the long-axis of the bar and is not aligned with the field in the neighboring region. The polarization in MSR4 may be dominated by the foreground cloud and unrelated to the bar, different than for the other three MSRs.

\section{Discussion} \label{sec:dis}

The Keyhole Nebula is a unique opportunity to probe the role of the magnetic field in a cloud strongly affected by stellar feedback. \citet{smith02} suggested that the Keyhole Nebula was previously a filamentary cloud nearby $\eta$ Carinae and that the stellar wind or outflow originating from $\eta$ Carinae pushed a part of the Keyhole Nebula, forming the loop in the middle of the filamentary structure. That the magnetic field is well-aligned with the bar and the loop, and that the magnetic field between the bar and the loop is in a fan shape, suggests that the magnetic fields may have been deformed together with the gas during the loop formation. 

The important question is whether the magnetic field plays an important role in the process, or whether it is frozen in the gas  and dragged along by the gas during the loop formation. To address this question, we carry out a simple calculation comparing the magnetic field tension with the ram pressure of stellar wind/outflow from $\eta$ Carinae.   

The ram pressure of the stellar wind/outflow can be written as
\begin{equation}
P_{ram} = \rho_w v_w^2,
\end{equation}
where $P_{ram}$ is the ram pressure, $\rho_w$ is the gas volume density of the wind at the loop, and $v_w$ is the wind velocity at the loop. The wind velocity has been estimated in multiple studies \citep[e.g.,][]{smith02,davidson18}, using optical and UV emission line profiles. While the exact wind speed is still uncertain, estimates of the terminal wind speed range from 300 km s$^{-1}$ to 1000 km s$^{-1}$. We assume that the wind speed at the Keyhole Nebula is the same as the terminal speed because the Keyhole Nebula is sufficiently distant from $\eta$ Carinae that the stellar wind is able to reach the terminal speed. The gas volume density was not determined in these studies since it cannot be estimated using the optical and UV lines that were used to estimate the velocity. \citet{oberst11} estimated the  density of the Tr 16 HII region using ISM models, finding $n_p$ = 20 cm$^{-3}$. We use this density as a lower limit to the wind density since the HII region density is the ambient density in the Tr 16 region. With these values and assumptions, the ram pressure is $\sim$3 $\times$ 10$^{-8}$ dyne if we take $v_w$ = 300 km s$^{-1}$, and $\sim$3 $\times$ 10$^{-7}$ dyne if we take $v_w$ = 1000 km s$^{-1}$.

The magnetic tension is  
\begin{equation}
f_{tension} = {B_0^2 \over R_c},
\end{equation}
where $f_{tension}$ is the force per unit volume exerted by the magnetic tension, $B_0$ is the magnetic field strength, and $R_c$ is the radius of curvature of the field. The magnetic field tension is the force per unit volume, while the ram pressure is the force per unit area. To make a comparison, we obtain the force per unit area from the magnetic tension by multiplying the force per unit volume by a characteristic dimension of the loop; thus $P_{tension} = f_{tension}\times l$, where $l$ is the thickness of the loop. From the H$\alpha$ image, we estimate the radius curvature as 0.6 pc, and the loop thickness as 0.08 pc. 

The magnetic field strength cannot be directly estimated from the polarization since the polarization is a result of multiple physical processes, including density fluctuations, turbulence, and coupling of the magnetic fields to gas and to dust grains. \citet{skalidis20} suggested that the magnetic field strength may be estimated as $B_0 = \sqrt{2\pi\rho}\delta v/\sqrt{\delta\theta}$, where $\rho$ is gas density, $\delta v$ is the gas $rms$ velocity, and $\delta\theta$ is the polarization angle dispersion, using several assumptions which are often adopted for a turbulent ISM. \citet{oberst11} estimated the gas density in the Keyhole Nebula through detailed PDR modeling. The gas density in the loop is roughly $n_H$ = 1000 cm$^{-3}$. We estimate the gas $rms$ velocity from the width of the HI 21 cm line \citep{rebolledo17}. The Keyhole Nebula feature appears as an absorption line at -30 km s$^{-1}$ with $rms$ velocity width of 3 km s$^{-1}$. The dispersion of the magnetic field angle is estimated within 5$''$ radius along the thick green line in Figure \ref{fig_structures}. We found that the local dispersion stays within $\sim$0.15 radians. Using these values, we estimate a magnetic field strength of $\sim$70 $\mu$G, which is of the same order as those found in the other star-forming regions \citep[e.g.,][]{koch12,koch18,soam19}.

Combining the above estimates, we calculate the pressure exerted by the magnetic field tension to be $\sim$4 $\times$ 10$^{-10}$ dyne. This is at least two orders of magnitude smaller than the ram pressure produced by the stellar wind from $\eta$ Carinae. This suggests that the magnetic field in the Keyhole Nebula is not strong enough to stop or maintain the structure against the impact of the stellar wind. The well-aligned magnetic fields in the loop may  simply be a result of the tight coupling of magnetic field to the gas during the formation of the loop by the stellar wind.

\section{Summary \& Conclusions} \label{sec:sum}

In this paper, we have presented the dust continuum map at 89 $\mu$m and the projected magnetic field orientation in the Keyhole Nebula region in the Carina Nebula Complex (CNC). We investigated the physical properties of the magnetic field in the Keyhole Nebula to understand the role of the magnetic field in a region influenced by strong stellar feedback. The main results are the following.

1. SOFIA HAWC+ has successfully observed the Keyhole Nebula and $\eta$ Carinae in the CNC with polarization uncertainty less than 0.5\%. The degree of polarization in the Keyhole Nebula has a mean value of 2.4\% with a standard deviation of 1.6\%. These values are similar to other high--mass star--forming regions such as OMC-1 and M17, which are much weaker in terms of star formation activity and stellar feedback than the CNC. This may indicate that the degree of polarization is not sensitive to the strength of stellar feedback.

2. The magnetic field orientation in the bar structure in the CNC is almost identical to that of the large-scale magnetic field, while that of the loop does not show any correlation with the large--scale magnetic field. We found that the orientation of the magnetic field in the region between the bar and the loop exhibits a fan shape starting from the bar and extending to the loop. This agrees with the picture in which the loop has formed due $\eta$ Carinae's stellar wind and that the magnetic field in the loop and the area between the loop and the bar has been highly perturbed and systematically deformed along with structure of the gas. 

3. The magnetic field orientation in the regions with foreground clouds forms a continuous connection to those in their background structures, the loop, and the bar. This suggests that the polarized field is likely dominated by that in the bar and the loop while the field is less ordered in the cold and dense foreground clouds. 

Finally, comparing the ram pressure of the $\eta$ Carinae's stellar wind to the pressure exerted by the magnetic field tension, we concluded that the magnetic field in the Keyhole Nebula is not strong enough to maintain the current structure against the impact of the stellar wind from $\eta$ Carinae.

\acknowledgments
This research was carried out in part at the Jet Propulsion Laboratory, California Institute of Technology, operated under a contract with the National Aeronautics and Space Administration. Some of the observations used in this study were made using the NASA/DLR Stratospheric Observatory for Infrared Astronomy (SOFIA). SOFIA is jointly operated by the Universities Space Research Association, Inc. (USRA), under NASA contract NAS2-97001, and the Deutsches SOFIA Institut (DSI) under DLR contract 50 OK 0901 to the University of Stuttgart. Financial support for this work was provided to CDD and YMS by USRA through the Guest Observer Cycle 7 campaign (Proposal ID 07$\_$0081, PI: C. D. Dowell).

%% To help institutions obtain information on the effectiveness of their 
%% telescopes the AAS Journals has created a group of keywords for telescope 
%% facilities.
%
%% Following the acknowledgments section, use the following syntax and the
%% \facility{} or \facilities{} macros to list the keywords of facilities used 
%% in the research for the paper.  Each keyword is check against the master 
%% list during copy editing.  Individual instruments can be provided in 
%% parentheses, after the keyword, but they are not verified.

\vspace{5mm}

\bibliography{Keyhole_ref}{}
\bibliographystyle{aasjournal}

%% This command is needed to show the entire author+affiliation list when
%% the collaboration and author truncation commands are used.  It has to
%% go at the end of the manuscript.
%\allauthors

%% Include this line if you are using the \added, \replaced, \deleted
%% commands to see a summary list of all changes at the end of the article.
%\listofchanges

\end{document}